%% file: talk.tex
\documentclass[multphys,vecphys]{svmult}

\usepackage{makeidx}         
\usepackage{graphicx}        
\usepackage{multicol}        
\usepackage[bottom]{footmisc}

\makeindex             

\begin{document}

\title*{Dynamical Masses of Young Star Clusters: Constraints on the Stellar IMF and Star-Formation Efficiency}
\titlerunning{IMF and SFE Constraints in Clusters}
\author{Nate Bastian\inst{1} and Simon P. Goodwin\inst{2}}
\institute{$^1$ University College London
\texttt{bastian@star.ucl.ac.uk}\\
$^2$ University of Sheffield
\texttt{S.Goodwin@sheffield.ac.uk}}
%
%
\maketitle


\section{The Stellar Initial Mass Function in Clusters}
\label{sec:1}

Many recent works have attempted to constrain the stellar initial mass function (IMF) inside massive clusters by comparing their dynamical mass estimates (found through measuring the velocity dispersion and effective radius) to the measured light.  These studies have come to different conclusions, with some claiming standard Kroupa-type \cite{kroupa} IMFs (e.g. \cite{maraston}, \cite{larsen06}) while others have claimed extreme non-standard IMFs (e.g. the top or bottom of the IMF is over-populated with respect to a Kroupa IMF \cite{smith}).  However, the results appear to be correlated with the age of the clusters, as older clusters ($>$80~Myr) all appear to be well fit by a Kroupa-type IMF whereas younger clusters display significant scatter in their best fitting IMF \cite{bastian06a}.  This has led to the suggestion that the younger clusters are out of Virial equilibrium, thus undercutting the fundamental assumption which is necessary to derive dynamical masses.  We will return to this point in \S~\ref{sec:2} and \S~\ref{sec:3}.  Focusing on the older clusters, we see that they all have standard IMFs (see Fig~2), arguing that at least in massive clusters the IMF does not vary significantly.

\index{paragraph}
%
%
%
%
%
%
%

\begin{figure}

\includegraphics[height=8cm]{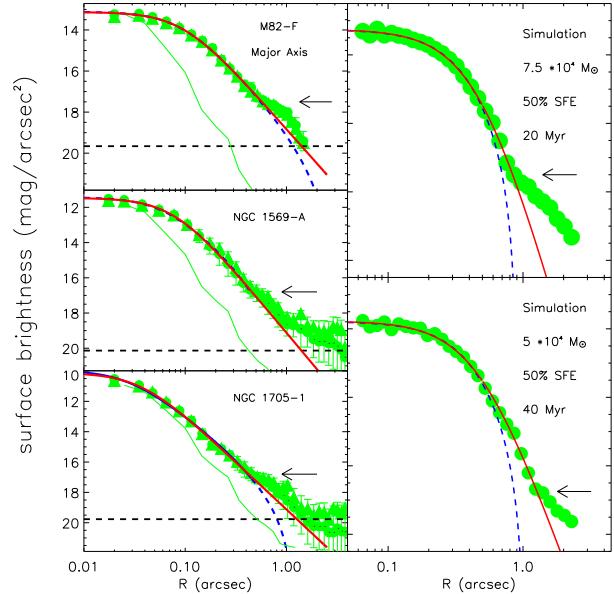}
\caption{{\bf Taken from \cite{bastian06b}:} Surface brightness profiles for three young clusters (left - M82-F, NGC~1569-A, and NGC~1705-1) and two N-body simulations which include the rapid removal of gas which was left over from a non-100\% star-formation efficiency (right).  The solid (red) and dashed (blue) lines are the best fitting EFF~\cite{eff} and King~\cite{king} profiles respectively.  Note the excess of light at large radii with respect to the best fitting EFF profile in both the observations and models.  This excess light is due to an unbound expanding halo of stars caused by the rapid ejection of the remaining gas after the cluster forms.  {\it Hence, excess light at large radii strongly implies that these clusters are not in dynamical equilibrium.} For details of the modelling and observations see \cite{bastian06b,goodwin}.}
\label{fig:2}       
\end{figure}

\begin{figure}
\includegraphics[height=9cm]{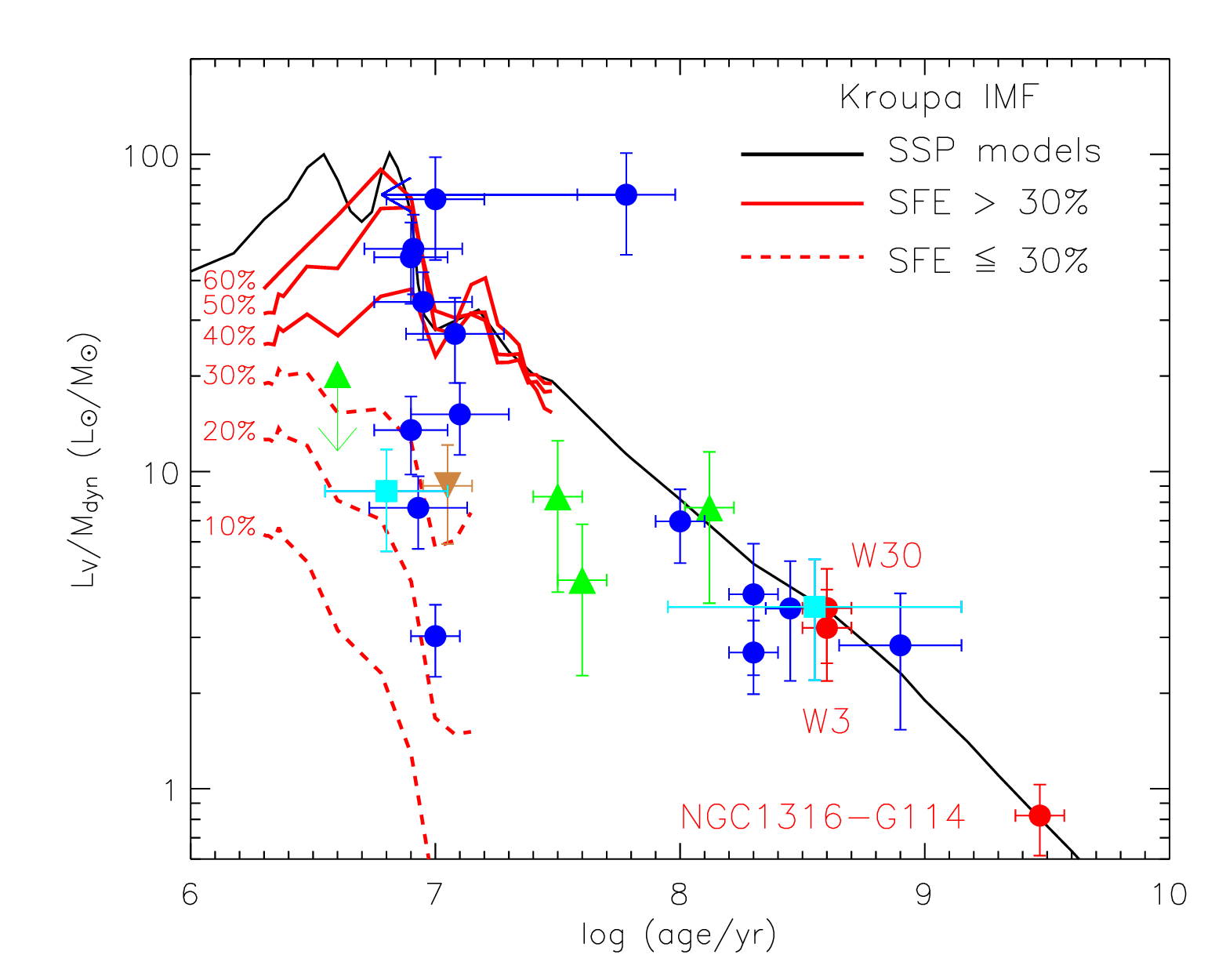}
%
%
\caption{{\bf Taken from \cite{goodwin}:} The light-to-mass ratio of young clusters.  The circles (blue and red) are taken from \cite{bastian06a} and \cite{maraston} and references therein, the triangles with errors (green) are LMC clusters \cite{mclaughlin}, the upside down triangle (brown) is for NGC~6946-1447 corrected for internal extinction \cite{larsen06}, and the squares (cyan) are from \cite{ostlin}. The arrow extending from M82F \cite{smith} is a possible correction to its age (see \cite{bastian06a}).  The triangle without errors is the tentative upper limit for cluster R136 in 30~Dor \cite{bosch,hunter}. The solid (black) line is the prediction of simple stellar population models (SSPs) with a Kroupa \cite{kroupa} stellar IMF. The red lines are the SSP model tracks folded with the effects of rapid gas removal following non-100\% star-formation efficiencies (SFE) \cite{bastian06b}. Dashed lines represent the SFEs where the clusters will become completely unbound. The SFE in the simulations
measures the degree to which the cluster is out-of-virial equilibirum
after gas loss, and so is an {\em effective} SFE (see \cite{bastian06b,goodwin}).}
\label{fig:1}       
\end{figure}

\section{Dynamical Equilibrium of Young Clusters}
\label{sec:2}

One explanation of why the youngest clusters are not in dynamical equilibrium is that young clusters are expected to expel their remaining gas (left over from the star-formation process) on extremely rapid timescales, which will leave the cluster severely out of equilibrium (e.g.~\cite{goodwin97a}).  In order to search for such an effect we compared the luminosity profiles of three young clusters with that of N-body simulations of clusters which are undergoing violent relaxation due to rapid gas loss \cite{bastian06b}.  The simulations (Fig~1, right panel) make the generic prediction of excess light at large radii (with respect to the best fitting EFF profile \cite{eff}), due to an unbound expanding halo of stars which stays associated with the cluster for $\sim20-50$~Myr.  These stars are unbound due to the rapid decrease of potential energy as the gas is removed on timescales shorter than a crossing time (e.g.~\cite{goodwin97a}).  Observations of the three young clusters also show excess light at large radii (Fig.~1, left panel), strongly suggesting that they are experiencing violent relaxation \cite{bastian06b}.  Hence these clusters are not in dynamical equilibrium.

\section{The Star Formation Efficiency and Infant Mortality}
\label{sec:3}

Assuming that young clusters are out of equilibrium due to rapid gas loss (the extent of which is determined by the star-formation efficiency - SFE one can fold these effects (see Fig.~3 in \cite{bastian06b}) into SSP models \cite{goodwin}.  The results are shown as solid and dashed red lines in Fig.~2 for various SFEs, where we have assumed all gas is lost instantaneously at 2~Myr.  The dashed lines show the results for SFEs below 30\% for which the cluster will become completely unbound.  Solid lines represent SFEs above 30\% where a bound core may remain.  Note that the observed SFEs of the clusters range from 10-60\% \cite{goodwin}.  

We also note that 7 out of the 12 clusters with ages below 20~Myr appear unbound (i.e. SFE~$<$~30\%), suggesting that $\sim60$\% of clusters will become unbound in the first 20-50~Myr of their lives \cite{goodwin}, i.e.~what has been termed ``infant mortality''.  This is in close agreement with cluster population studies of M51 which found an infant mortality rate of 68\% \cite{bastian05} and comparable to the open cluster dispersal rate of $\sim87$\% \cite{lada} (see also \cite{whitmore03}).

\section{Conclusions}

Through detailed comparisons of the luminosity profiles of young clusters with N-body simulations of clusters including the effects of rapid gas loss, we argue that young clusters are not in Virial equilibrium.  This undercuts the fundamental assumption needed to determine dynamical masses.  This suggests that the claimed IMF variations are probably due to the internal dynamics of the clusters and not related to the IMF.  By limiting the sample to the oldest clusters (which appear to be in equilibrium) we see that they are all well fit by a Kroupa-type IMF arguing that, at least in massive star clusters, the IMF does not vary significantly.

By combining the above N-body simulations with SSP models we can derive the (effective) SFE of clusters.  From this we find that $\sim60$\% of young clusters appear to be unbound, in good agreement with other estimates of the infant mortality rate.  Note however that even if a cluster survives this phase it may not survive indefinitely due to internal and external effects (e.g.~\cite{gieles}).

\begin{acknowledgement}
NB gratefully thanks his collaborators Roberto Saglia, Paul Goudfrooij, Markus Kissler-Patig, Claudia Maraston, Francois Schweizer, and Manuela Zoccali on dynamical mass studies.
\end{acknowledgement}
%
%
%
\input{referenctalk}



\printindex
\end{document}

%% file: referenctalk.tex
%
%

%
%